# Origin and Enhancement of Spin Polarization Current in Diluted Magnetic Oxides by Oxygen Vacancies and Nano Grain Size

Hsiung Chou[1, *], Kung-Shang Yang[1], Yao-Chung Tsao[1], G. D. Dwivedi[1], Cheng-Pang Lin[1], Shih-Jye Sun[2], L. K. Lin[3], S. F. Lee[3]

[1]*Department of Physics, National Sun Yat-Sen University, Kaohsiung 804, Taiwan*

[2]*Department of Applied Physics, National Kaohsiung University, Kaohsiung, Taiwan*

[3]*Institute of Physics, Academia Sinica, Taipei, Taiwan*

## Abstract

Spin polarized current is the main ingredient of diluted magnetic oxides due to its potential manipulation in spintronic devices. However, most research has focused on ferromagnetic properties rather than polarization of electric current, because direct measurements were difficult and the origin of spin polarized currents has yet to be fully understood. The method to increase the spin polarized current percentage is beyond practical consideration at the present status. To target this problem, we focus on the role of oxygen vacancies and nano grain size on the spin polarized current, which are controlled by growing the Co-doped ZnO thin-films at room temperature in a reducing atmosphere [Ar + (1% ~ 30%) $H_2$]. We found that the conductivity increases with an increase in oxygen vacancies via two independent channels: the variable range hopping (VRH) within localized states and the itinerant transport in the conduction band. The Andreev Reflection measurements prove that the VRH conduction is equivalent to spin polarized current. Transport measurements show that the best way to increase the VRH content, or the percentage of spin polarized current, is to increase oxygen vacancy concentration and to reduce grain size and lattice constants of the films.

*Corresponding Author: hchou@mail.nsysu.edu.tw

**Introduction:**

Ever since its discovery, diluted magnetic oxides have attracted the attention of many scientists and engineers who aim to understand and apply them to spintronic devices. The most significant feature of diluted magnetic oxides is the spin polarized current (SPC), the crucial element for their application [1-4], especially for those oxides which have the potential to work above room temperature. To date, only very few reports mention an observation of spin polarized current in diluted magnetic oxides (DMOs) through either Andreev reflection [5] or tunneling magnetoresistance [6,7] measurements. In order to be used in room temperature devices, it is necessary to determine the origin of this spin polarized current as well as the mechanism of ferromagnetic coupling in diluted magnetic oxides; as such, many mechanisms have been proposed by researchers [8-21]. It has been found that the ferromagnetic coupling can occur in metallic [5], semiconducting [22-24], and nearly insulating [8] states. Gehring *et al*. [5] used Andreev reflection to study the spin polarized current in ZnO films co-doped with Mn and Al in the metallic state and found a spin polarization of about 60%. They concluded that the room temperature ferromagnetic coupling and the spin polarization properties were the effects of Mn doping, as these features were not observed in Al doped ZnO samples. Xu *et al*. [6] used high resistance Co-doped ZnO as a spin filter sandwiched between an Al doped ZnO bottom layer and a metal Co layer. Their results support the existence of spin polarization current without presenting any explanation of the underlying physics. In order to extend its distinct characteristics to room temperature applications, it is crucial to find the origin and formation details of the spin polarized current and then develop the method to control it.

In this manuscript, we have used transport measurement and Andreev reflection to clarify the origin of spin polarized current on Co-doped ZnO thin-films. We found that the variable range hopping (VRH) channel is responsible for the spin polarized current while the thermal excitation (TE) channel is not. With proper control of grain sizes, lattice constants and oxygen vacancies in the present thin-films, we were able to increase the percentage of VRH channel up to 80% at room temperature which equals to 80% spin polarized current.

**Experimental Details:**

A series of $Zn_{0.95}Co_{0.05}O_{1-\delta}$ (CZO) thin-films were grown, by standard magnetron sputtering in a reducing atmosphere, on glass and fused quartz substrates. For reducing atmosphere, we used a mixture of Ar and 1% ~ 30% of $H_2$ (specified as #% such that CZO films grown at a specific $H_2$% are written as CZO-#%). Unlike other high temperature processing conditions, these thin-films were grown at room temperature in order to reduce the possibility of hydrogen doping in films.

Crystal structures and phase purity of as-grown thin-films were examined by a Bede D1 High resolution X-ray diffractometer. Near edge X-ray absorption spectroscopy (NEXAS) and extended X-ray absorption fine structure (EXAFS) were performed in Taiwan's NSRRC 01C1 beam line, on the Co K-edge. Optical transmittance spectra were recorded with an N&K analyzer 1280 (N&K technology Inc.). Conductivity measurements were performed via a conventional four probe method. The conductance-voltage (G-V) curves were measured at 4.2 K temperature by the differential technique. A three-step delta technique was used to measure the differential conductance as a function of applied voltage on the point contacts. We used a Keithley 6221 current source to provide an alternating staircase sweep current. The amplitude of the alternating portion of the current is the differential current (dI). Each delta voltage is averaged with the previous delta voltage to calculate the differential voltage (dV). The differential conductance ($\Delta G$) can be derived using dI/dV. Each conductance curve was normalized by its high-bias value for further analysis.

**Results and discussion:**

Since Co-clusters and incorporated hydrogen in ZnO crystal lattice can disturb the spin polarized current and the ferromagnetic property of DMO due to Co-clusters' magnetic moments or H shallow donor states, it is obvious to ensure that the thin-films had been grown without any Co-clusters and H incorporation. The crystal structures and phase purity of the as-grown nanocrystalline thin-films are examined by a grazing angle incidence X-ray diffraction (GIXRD). X-ray diffraction patterns of polycrystalline thin-films grown with various percentages of forming gas $H_2$ are plotted in Figure 1. Diffraction patterns of CZO films are in agreement with Bragg's peaks for hexagonal wurtzite structures with space group *P6₃mc*. The (00$\bar{2}$) peak exhibits the highest intensity, indicating a preferred-orientation of the CZO crystals. When $H_2$% becomes high enough, a certain amount of oxygen in plasma reduces into water and generates numerous oxygen vacancies in the CZO crystal. In contrast, no cobalt metal clusters are observed by XRD. By increasing $H_2$%, the ZnO ($00\bar{2}$) peak shifts toward higher angle, indicating a smaller *c*-axis lattice constant, and its full widths at half maximum (FWHM) is found to increase, which suggests shrinking of crystal size. Lattice constant '*c*' and grain sizes of all hexagonal structured CZO thin films are calculated with the most intense ($00\bar{2}$) peak. We found that lattice constant '*c*' and grain size decreases from 0.525 nm and 25 nm for CZO-1% to 0.522 nm and 14 nm for CZO-30% samples, as shown in the inset of Figure 1. Incorporation of hydrogen at the grain boundary surface as well as at the crystal lattice of ZnO films can be easily accomplished in four ways: by post-growth annealing at high temperature in an $H_2$ environment [25], by growing films at high temperature in a mixed forming gas which contains $H_2$ [26-28], by implantation [29] or by plasma treatment [30]. However, the easy diffusion of hydrogen into a lattice is also countered by its effusion at high temperatures [31]. Such temperatures are critical for most multilayer processing, resulting in a severe deterioration in physical properties. This strongly implies that precise control of hydrogen content in films is very difficult. If H incorporates into the material as a shallow donor, the increased electrical conductivity will be a combined effect of the huge electron transport in the conduction band by thermal excitation from the both hydrogen shallow donor states and the oxygen vacancy states. With uncontrollable interstitial hydrogen content at each step of processing, the conduction situation will be very complicated. In order to avoid this problem, it is necessary to develop a growth procedure that can constrain the role of $H_2$ in the forming gas to only generating oxygen vacancies while not allowing $H_2$ to incorporate into the ZnO crystal lattice. This situation can be accomplished simply by growing films at room temperature. If H incorporates in the interstitial sites, such as the bond center sites perpendicular or parallel to the *c*-axis, or in the antibonding sites, the $d_{002}$ will increase [26]. If hydrogen occupies the oxygen vacancy sites, $H_O$, the $d_{002}$ may decrease, accompanied by larger crystal size [26]. In the present samples, however, since both $d_{002}$ and crystal size decrease, it is highly likely that hydrogen does not incorporate in any of the ways mentioned above inside the CZO crystal.

To investigate the possible existence of metal Co clusters that cannot be detected by GIXRD, NEXAS and EXAFS were performed on the Co K-edge of CZO-10%, -20% and -30% samples because they can contain high levels of oxygen vacancies that may facilitate the segregation of Co dopants to form a higher level of Co clusters. We have also performed NEXAS and EXAFS for pure Co metal, CoO and Co-0.06%:ZnO film as references (Figure 2). The Co ion level in the Co-0.06%:ZnO film is much lower than the solubility of Co (~3.75%) [23] in the CZO thin films grown by the δ-technique and must be fully dissolved in the ZnO matrix. The pre-edge of 1s to 3d transition around 7709 eV is the main characteristic distinguishing the Co clusters, CoO and CZO films because the Co K-edge spectra is quite different for the Co metal, CoO and CZO films. In Figure 2, a clear hump within the 7709~7715 eV range for Co metal film is observed, while no hump is observed for the Co-0.06%:ZnO film in that region. In other words, this indicates that the Co K-edge spectrum for any CZO samples without Co-cluster precipitations should follow that of the Co-0.06%:ZnO curve. NEXAS spectra of CZO-10%, -20% and -30% samples completely follow that of the Co-0.06%:ZnO film, with no evidence of Co clusters within the experimental resolution. The radial distribution of Co calculated from EXAFS, in the inset of Figure 2, shows that the first and second nearest neighbor patterns of present films are similar to the Zn pattern for pure ZnO, yet totally different from the Co pattern from Co metal and CoO. The average distances of the central Co ion to its first and second nearest neighbors are 1.84 and 3.13 Å, while the distances of the Zn ion to its first and second nearest neighbors in a pure ZnO crystal are 1.74 and 3.16 Å. The slightly shorter Co-second nearest neighbor distance, when compared to Zn, is a direct reflection of the smaller ionic radius (0.65 Å) of the Co ion when compared to the Zn ion (0.74 Å). From these data, we believe all doped Co ions are incorporated into the Zn site, therefore no Co-clusters are found by GIXRD. Note that even though the doping concentration of Co ions in the present films is higher than the solubility for films used in previous studies [23], no Co clusters are formed, which indicates that the solubility of Co is strongly dependent on the growth method.

Optical transmittance spectra of glass substrate (dashed black), CZO-1% (red), CZO-2% (orange), CZO-5% (dark yellow), CZO-10% (black), CZO-20% (green) and CZO-30% (blue) have been shown in Figure 3. It is observed that transmittance decreases with increasing $H_2$%, or in other words, with increasing oxygen vacancies. Optical transmittance spectra provide corroborative support for the non-existence of Co clusters by the appearance of three distinct characteristic absorption-bands of *d-d* transition of Co ions from $^4A_2(F) \rightarrow {}^2A_1(G)$, $^4T_1(P)$, and $^2E(G)$ at around 567, 613 and 662 nm for all samples, indicating the substitution of Co at Zn sites [32, 33]. Clearly, Co ions are dissolved in the ZnO matrix even when grown at such high oxygen vacancy concentrations. Moreover, spectra show two kinds of transition for each film, which have been highlighted with two boxes. First box shows a transition due to the absorption edge of the glass substrate, and second box represents the main transition from the film itself. The main transition provides the information about the optical bandgaps of thin films, which increase linearly with increasing oxygen vacancies (Inset Figure 3), while grain size decreases exponentially with increasing oxygen vacancies. This may be due to the Burstein-Moss blue shift effect, [34,35] or a combination of small size and short structural ordering effect. The latter explanation is better supported in our samples because the carrier concentration is in the range of $10^{19\sim20}$ cm$^{-3}$. Conclusively, the present CZO thin-films have no Co-clusters and H incorporations, the spin polarized current can be simply related to the intrinsic properties of the present films. In addition, the oxygen vacancy and band gap increases, while grain size and lattice constant decrease.

Conductivity in all films is found to increase with increasing oxygen vacancies. Since the ferromagnetism appears when $H_2 > 5\%$, in this manuscript, we focus on the thin-films with $H_2 > 5\%$. It has been reported in our previous study on the films grown by δ-growth technique [24] that the electrical conduction can be carried by two channels. To investigate whether the present thin-films exhibit a similar behavior, conductivities of present thin-films have been fitted by a combination of the thermal excitation (TE) and variable range hopping (VRH) models, $\sigma(T) = A\exp[-(C/T)^{1/4}] + B\exp(-E_d/kT)$, where A and B are constants, $E_d$ is the activation energy of the energy gap between the localized states and the lowest point of the conduction band, and C is the constant dependent on the hopping radius of carriers that are localized, as shown in Figure 4. The fitting confidence of samples with high conductivity (CZO-10% and -30% samples) is as high as 95%, while that of the low conductive sample (CZO-5%) is around 90%. The contribution of VRH and TE mechanisms to the conductivity are defined as $A\exp[-(C/T)^{1/4}]/\sigma(T)$ and $B\exp(-E_d/kT)/\sigma(T)$, respectively, and are plotted in the inset Figure 4(a). It is clear that the TE mechanism drops off very quickly at lower temperatures where the VRH mechanism dominates the conductivity. At room temperature, the contribution of the VRH mechanism is enhanced dramatically from 5% of the CZO-5% sample to 80% for the CZO-30% sample (inset Figure 4(b)). The enhancement on VRH mechanism indicates both the increasing oxygen vacancies and reducing the grain size and lattice constants show profound effect on conduction behavior.

Andreev reflection (AR) is the best tool to investigate the contribution of hopping conduction in inducing spin polarized current. Andreev reflection observes the carrier generation, or reflection, on the metal side of the superconductor/metal interface, and the formation probability of reflection carriers is highly dependent on the spin polarization of conduction current of the measured sample. The enhancement of AR conductivity will be observed for normal metals containing similar density of spin states on both majority and minority bands around the Fermi surface. In contrary, the suppressed conductivity will be measured for ferromagnetic conductors with unequal density of spin states on both majority and minority bands around Fermi surface. Ideally, electric conductivity becomes zero for a half metal or 100% polarized ferromagnetic materials.

In this experiment, a 99.9% pure Pb wire of 2 mm diameter was sharpened by a ceramic scissor into a nano-scale sharp tip. The Pb tip was mounted on a differential screw stage for precise control of perpendicular movements toward the surface of the present films. The whole assembly was kept in a liquid He bath to cool the setup to 4.2 K, where the Pb-tip will be in a superconductor state. At 4.2 K, where the conduction of all films is entirely dominated by the variable range hopping mechanism, the Andreev reflection measurement condition is best suited to our purpose of investigating the spin polarization from the hopping conduction. The superconducting tip was moved carefully to make electrical contact with the surface of the film. Contact was confirmed whenever the G-V measurements presented a non-zero reading. To minimize the thermal effect at the contact point, a differential measuring technique [36] was applied. An effective G-V curve is defined as fulfilling the following two criteria: (1) the G-V curve must be narrow because the superconducting gap for Pb metal is only 1.35 meV. Any possible effect that broadens the G-V curve must be in reasonable range; and (2) the contact resistance must be relatively large in series contact tests to meet criteria of the ballistic tunneling limit for a small point contact. In Figure 5, the green and purple curves with contact resistances of 1.72 and 1.64 kΩ indicate bad contacts because they had turning points larger than 10meV and at the high bias region, where the tunneling was made into the normal state of the Pb tip, the curve was not flat as expected but instead keeps increasing with bias voltages or exhibits large noise values. We continuously changed contact points until both these criteria were satisfied, as shown as the red curve in Figure 5. At this stage, the contact point might not be perfect, as the curve was fitted to have a large contact barrier, the Z values. Then the Pb tip was pushed deeper into the film. Subsequent G-V curves were measured following every pushing movement until the turning points of a curve were widened again, and displayed large noise values and unstable and non-horizontal curves at high bias voltages, shown as the black and blue curves. The blue curve of Figure 5 corresponds to an over contact situation, where the tip is flattened and the contact area was enlarged, providing more conduction channels and inelastic scattering at the interface such that both the width and noise increased.

The present samples contain various oxygen vacancies and displayed very different conductivities. Their best curves under contact conditions are shown in Figure 6. The CZO-5% and -10% samples have very high contact resistances with Pb tips at 4.2 K, as high as 46 k$\Omega$ and 16.7 k$\Omega$, respectively, indicating a very tiny contact area. As a result of this, the dips in the G-V curves for the CZO-5% and -10% samples are relatively wide. The full width at half height (FWHH) of G-V curves for all samples are shown in the inset and decreases sharply when the oxygen vacancies increase. The CZO-20% and -30% samples manifest the narrowest dips in the G-V curve. To ensure a reasonable and reliable reading, we focus only on the CZO-30% sample, as it has better G-V curves.

The original Blonder-Tinkham-Klapwijk (BTK) model [37] considered tunneling between normal-metal/superconductor when the contact area is smaller than the mean free path of electrons. Mazin *et al.* [38] modified the BTK model (as the MBTK model) to extend it to ferromagnetic metal, and to include both ballistic and diffusive limits. The MBTK model includes the parameters of barrier height (Z), the spin polarization of the ferromagnetic material (P), and the superconductor gap ($\Delta$). The resistances of the present films are relatively high, hence whenever the current flows into the small contact point, a huge Joule heating effect is generated before and after tunneling to broaden the G-V curve. Two additional effects of spreading resistance [39] and effective temperature [40] were vital for simulation and fitting. To reduce the uncertainty of the fitted parameters and the unreasonable spin polarization ratio due to fitting too many parameters simultaneously, we set the superconducting band gap at a fixed 1.35 meV. Under this constraint, we could fit the experimental curves with the models and the systematical change in the spreading resistance and effective temperature was observed. Therefore, the spin polarization of the present Co-doped ZnO with various contact barriers could be resolved.

To illustrate the fitting processes and considerations, Figure 7 presents the G-V curve of the CZO-30% sample fitted with both ballistic, (a), and diffusive, (b), limits. The unsymmetrical G-V curves at positive and negative bias region were due to the different shape of the Pb tip electrode and the FM film. The rest of the G-V curves were fitted by the same procedure to resolve the spin polarizations and Z values at various contact conditions. According to Figure 7, both limits provide an equally good fit in the smaller bias region, within the turning points demarcated with green triangles. However, at the larger bias regions, outside the green triangles, the ballistic model exhibits better fit than that of the diffusive model. Around the turning point, the data curve did not show any peak with the high Z value indicating that our contact point is free from oxides that might cause severe scattering.

The simulation results of the G-V curves on Z and polarization, P, is plotted in Figure 8. Both limits show a linear relationship, as plotted with the dashed guide lines. The ballistic limit gives rise to a higher polarization than that of the diffusive limit. To date, there have been no absolute extrapolation processes, either quadratic, linear or spin flip dependent approaches, that can be used to describe the P-Z relation of all systems [41-44]. Some reports have mentioned that the high resistance contact will increase the difficulty in correctly estimating spin polarization [45-47]. The present analysis process described above can certainly provide a reliable estimated polarization value. With Z values approaching zero, spin polarization ratios of 73% and 67% were estimated for the ballistic and diffusive limits, respectively. These results provide direct evidence of spin polarization current induced by hopping conduction.

If the hopping conduction leads to the spin polarization current, the polarization should be 100% in a perfect measurement condition. This discrepancy may be understood as the nature of the hopping mechanism. In the present model, the carriers were localized around oxygen vacancies forming VRH spheres with a single digit nanometer scale radius. Hopping conduction occurred only for those spheres which are situated within the hopping distance. The ferromagnetic phenomenon is achieved only when a percolation path exists through the entire sample from one end to the other. The percolation path consists of tiny VRH spheres, similar to a dotted line, with a very thin single digit nanometer scale width. When the Pb-tip makes contact with the film, the contact point may reach hundreds of nm in size and might or might not cover the percolation paths.  That is why the first contact usually did not show the correct G-V curves.  Making contact at various places was necessary to find the optimal contact for further measurements.  Since the contact area is much larger than the width of every percolation path, this contact area will likely cover both percolation paths as well as some fractures that do not run the length of the samples, and therefore act as superparamagnet.  Therefore, the measured spin polarization ratio will always smaller than 100%.

This study reveals that the hopping conduction is the backbone of spin polarization current in DMOs. This property will remain intact in all temperature range as long as the hopping conduction contributes to the electric conduction.  At room temperature, the spin polarization current is carried by the hopping conduction while the itinerant transport in the conduction band is a normal current.  Therefore, the spin polarization current is around 80% for the CZO-30% sample for it contains highest percentage of VRH conduction at room temperature. This finding reveals a route to enhance the spin polarization current at room temperature, i.e., by increasing oxygen vacancy concentrations.

**Conclusion:**

This study provided direct evidence that the hopping mechanism is the source of high spin polarization current, with oxygen vacancies playing an important role in providing localization sites for hopping.  Since oxygen vacancies may act as deep defect states or shallow donor states, only these oxygen vacancy states can overlap with the Co partial localized band or other possible localized magnetic states around the Fermi level to provide effective spin polarized hopping conduction. We found that an increase in oxygen vacancies and decrease in grain sizes can increase the percentage of the VRH mechanism at room temperature and hence can increase the spin polarization current ratio, which can prove beneficial for applications of room temperature spintronic devices.

**Acknowledgements**

The authors would like to acknowledge the great help from Professor Wei-Feng Tsai and Michael Chiang of National Sun Yat-sen University, Professor Cheng-Hsuan Chen of the Center for Condensed Matter Sciences, National Taiwan University, Taiwan, Professor Hideo Ohno and Fumihiro Matsukura of Tahoku University, Japan, Professor J. M. D. Coey of Trinity College, Ireland, and Professor Tomasz Dietl of Polish Academy of Sciences, Poland, for their great suggestions, questions and challenges. This project is supported by the Ministry of Science and Technology of Taiwan under Grant No. NSC-102-2112-M-110-MY3.

**Figure Captions:**

**Figure 1:** The grazing incidence X-ray diffraction patterns. All ZnO peaks can be identified. One peak around 43° corresponds to Zn (011) for films grown at high $H_2$%, indicating large amounts of Zn Cluster precipitation. No trace of Co related metal or oxides is observed. The c-axis and grain sizes decrease with increasing $H_2$%, as plotted in the inset.

**Figure 2:** Near edge X-ray absorption spectra of Co K-edge for the CZO-10%, -20% and -30% samples and the Co metal and ZnO film doped with 0.06% of Co as references. Inset is the radial distribution of Zn and Co calculated from extended X-ray absorption fine structure (EXAFS). The radial distributions of Co in all CZO samples coincides with the radial distribution of Zn in a pure ZnO crystal, indicating all doped Co ions are incorporated into the ZnO structure.

**Figure 3:** Optical transmittance of films grown with various $H_2$% working gas concentrations. Two rectangular boxes indicate background and main transitions which correspond to the absorption edges of the glass substrates and the optical bandgaps of films, respectively.

**Figure 4:** Electrical conductivity as a function of temperature. The light curves at the center of each data curve are the fitted results using the VRH and TE combination model. The contributions of VRH and TE mechanisms and the room temperature conductivities are plotted in the inset figure (a). The contribution of VRH mechanism and the conductivity increase with increasing oxygen vacancies. Inset figure (b) presents the contribution of VRH in room temperature conductivity for 5%, 10% and 30% $H_2$ samples.

**Figure 5:** Conductance vs voltage behavior of CZO-12% film shown at various contact resistances. As contact resistance increases, the G-V curve shows a sharp dip (decreased FWHH), which is an indication of spin polarization.

**Figure 6:** Conductance vs voltage curve for CZO-4%, CZO-6%, CZO-8% and CZO-12%. FWHH decreases with increasing oxygen vacancies, which demonstrates that oxygen vacancies can control spin polarization current. Inset figure shows the trend of FWHH with increasing oxygen vacancies.

**Figure 7:** Comparative study of conductance vs voltage curve of CZO-12% under two different models: (a) ballistic and (b) diffusive. This clearly shows that ballistic model can describe the curve more closely and provide higher spin polarization (70%) than the diffusive model (60%).

**Figure 8:** A linear relationship between barrier height (Z) and spin polarization (P) for CZO-12%.

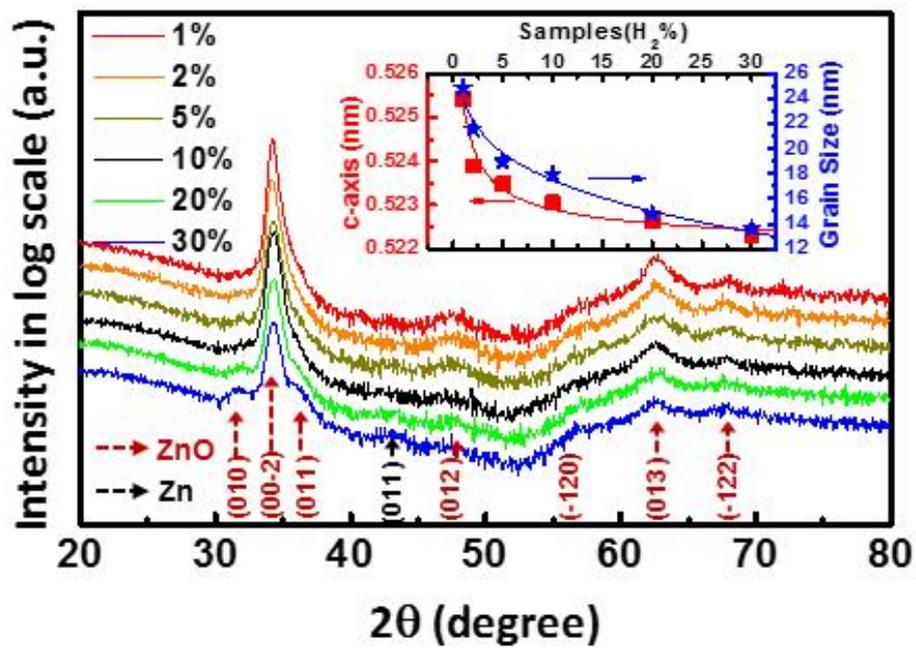

Figure 1

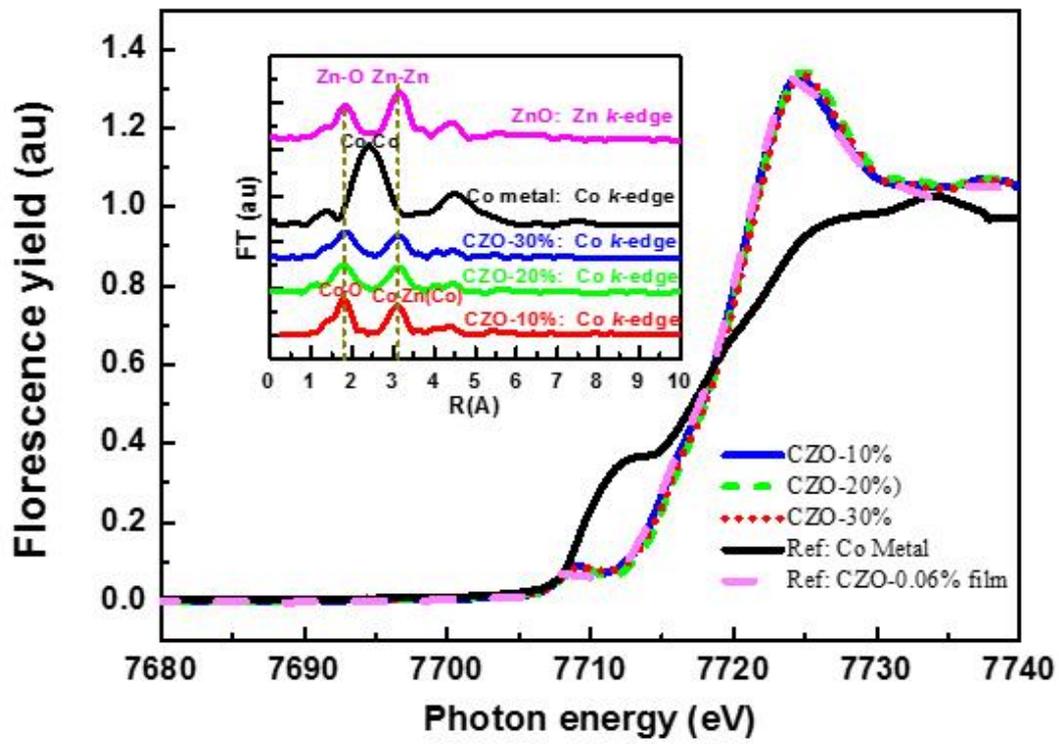

**Figure 2**

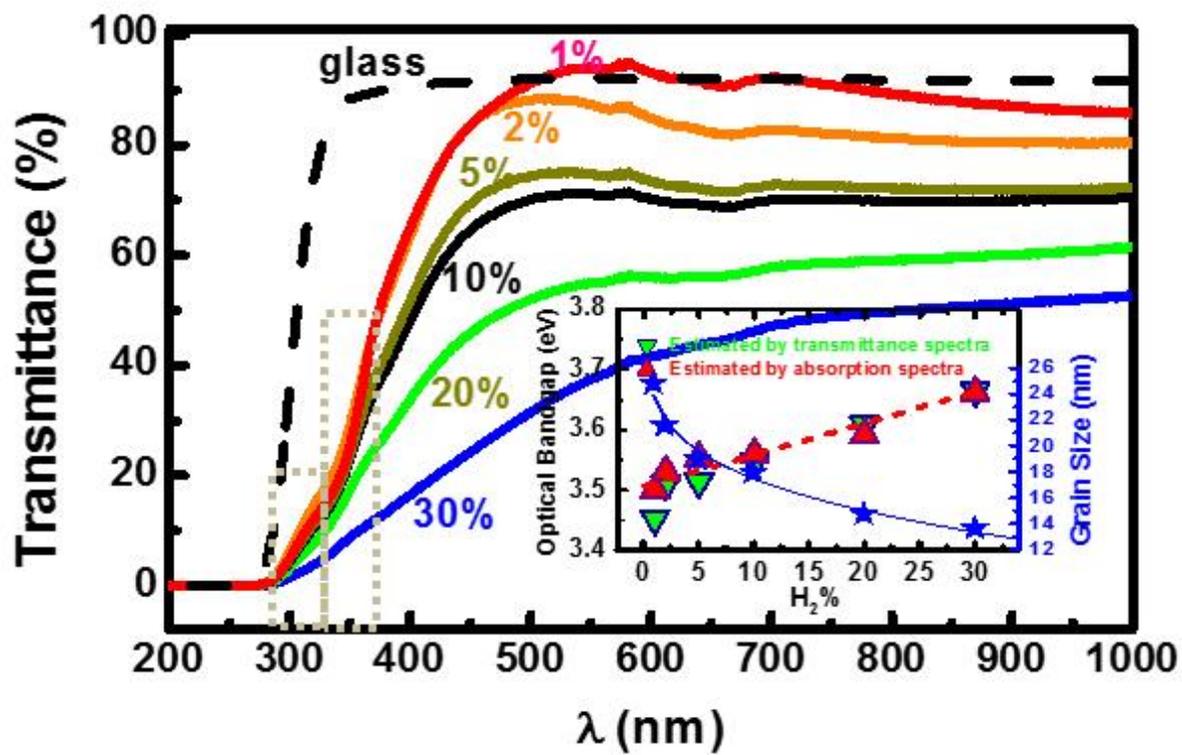

Fgiure 3

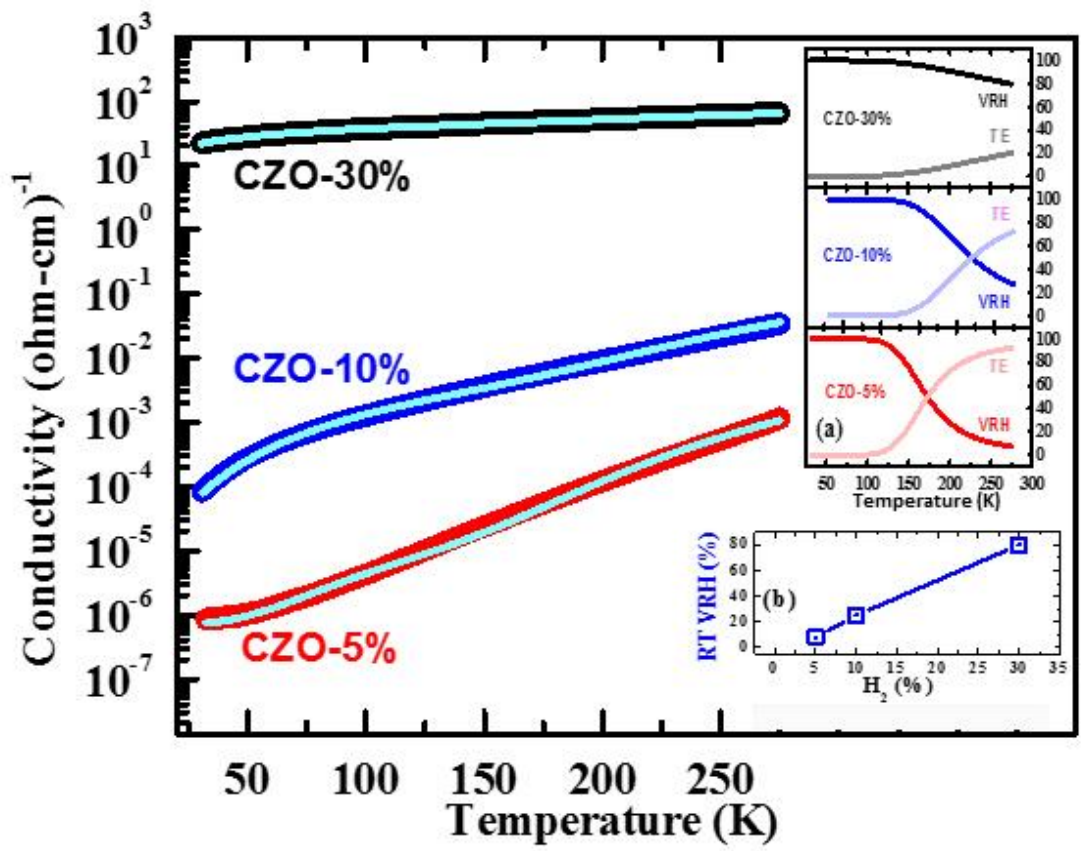

**Figure 4**

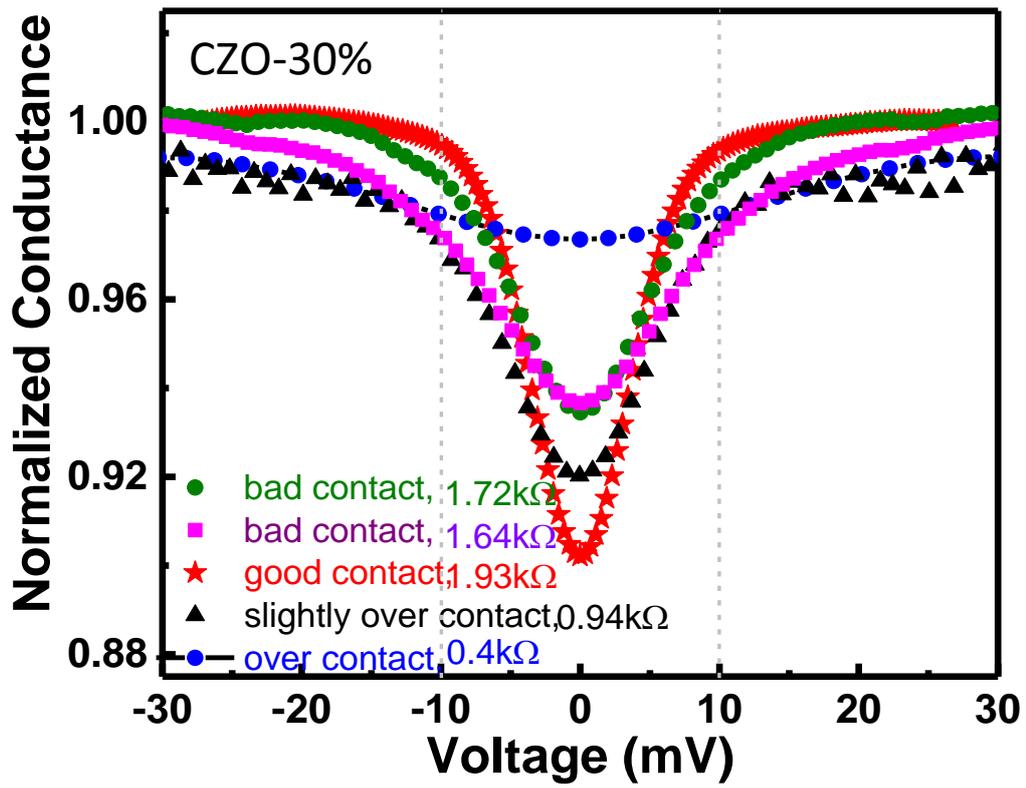

Figure 5

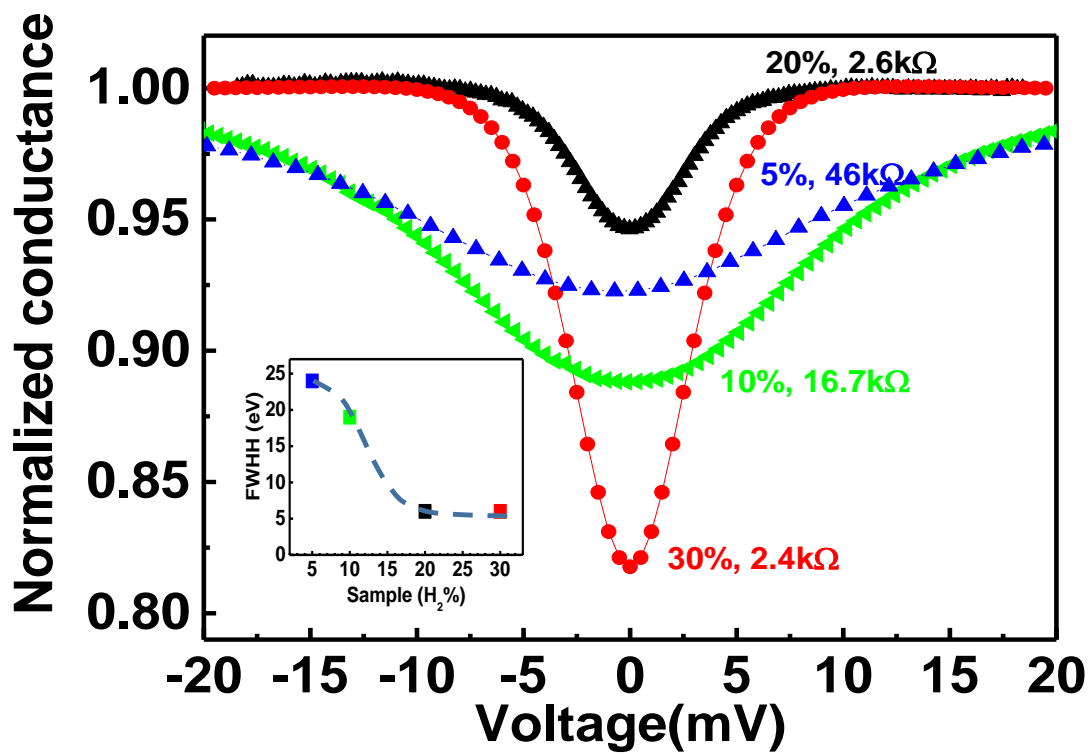

**Figure 6**

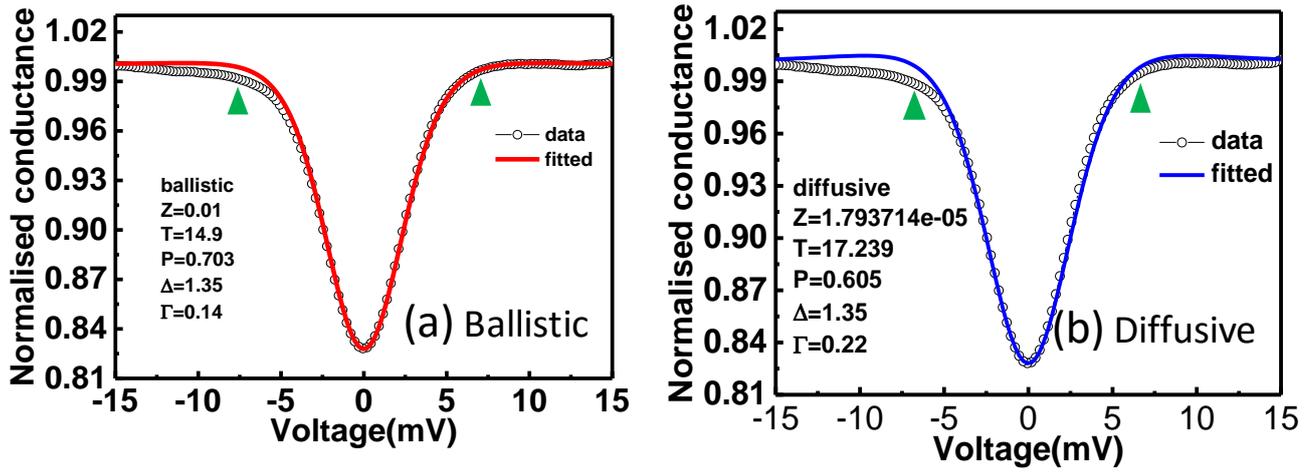

**Figure 7**

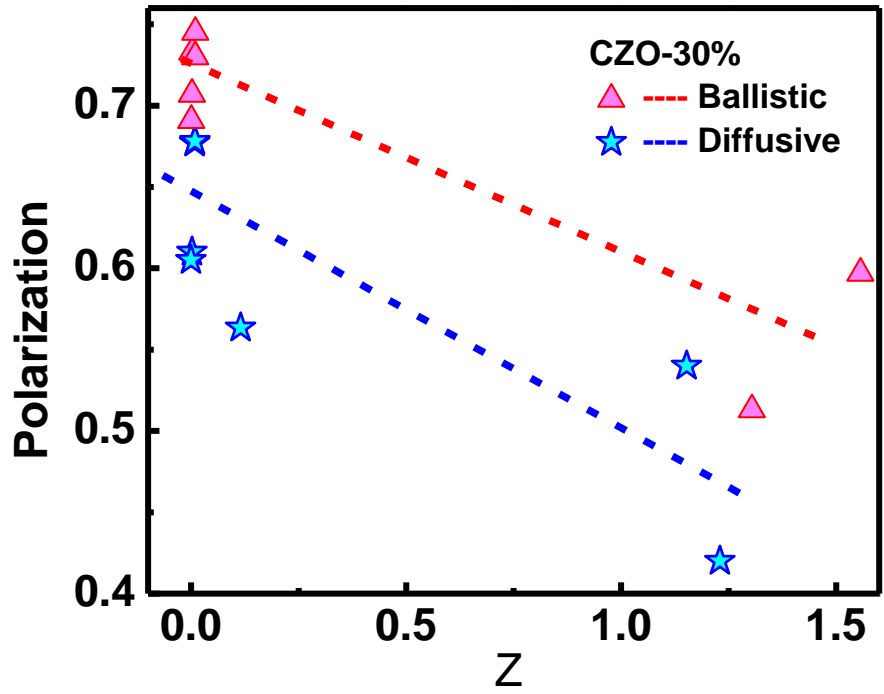

Figure 8